**Proximity in face-to-face interaction is associated with mobile phone communication**

Tobias Bornakke

University of Copenhagen

Talayeh Aledavood and Jari Saramäki

Aalto University

Sam G. B. Roberts

Liverpool John Moores University

Author Note

Tobias Bornakke Jørgensen, University of Copenhagen, Øster Farimagsgade 5A, Bld. 16, PO Box 2099, DK-1014, Copenhagen, Denmark; Talayeh Aledavood and Jari Saramäki, Department of Computer Science, Aalto University, PO Box 11000, FI-0076, Aalto, Finland; Sam G. B. Roberts, School of Natural Sciences and Psychology, Liverpool John Moores University, Liverpool, L3 3AF, UK

Tobias Bornakke Jørgensen is now at Analyse & Tal F.M.B.A. https://www.ogtal.dk/

We are very grateful to Professor David Dreyer Lassen and Dr Sune Lehmann for their support with this project. J.S. and T.A. acknowledge support from the Academy of Finland, Project Number 260427. S.R. is grateful to Dr Michelle Tytherleigh for helpful discussions relating to this project. This article was produced as part of the interdisciplinary 'Social Fabric' project, made possible through the support of the University of Copenhagen KU16 funding scheme. S.G.B.R. was funded by a research grant from the Department of Psychology, University of Chester.

Correspondence concerning this article should be addressed to Sam Roberts, School of Natural Sciences and Psychology, Liverpool John Moores University, Liverpool, L3 3AF, UK. Email: s.g.roberts1@ljmu.ac.uk

**Proximity in face-to-face interaction is associated with mobile phone communication**

## Abstract

The frequency of mobile communication is often used as an indicator of the strength of a tie between two individuals, but how mobile communication relates to other forms of 'behaving close' in social relationships is poorly understood. We used a unique multi-channel 10-month dataset from 510 participants to examine how the frequency of mobile communication was related to the frequency of face-to-face interaction, as measured by Bluetooth scans between the participants' mobile phones. The number of phone calls between a dyad was significantly related to the number of face-to-face interactions. Physical proximity during face-to-face interactions was the single strongest predictor of the number of phone calls. Overall, 36% of variance in phone calls could be explained by face-to-face interactions and the control variables. Our results suggest that the amount of mobile communication between a dyad is a useful but noisy measure of tie strength with some significant limitations.

**Proximity in face-to-face interaction is associated with mobile phone communication**

## Introduction

Human beings are above all social animals (Dunbar, 1998) and having strong and supportive relationships is essential both for psychological health and physical well-being (Baumeister & Leary, 1995; Holt-Lunstad, Smith, & Layton, 2010). The social relationships that people maintain with others vary widely, from very close, emotionally intense relationships to more casual acquaintances (Binder, Roberts, & Sutcliffe, 2012; Granovetter, 1973; Wellman & Wortley, 1990). Despite the widespread agreement that there are different types of social relationships, there is still no consensus on the most reliable way of measuring the nature of these relationships over time or in large samples.  Social psychological research has identified two distinct components of interpersonal closeness in relationships – *feeling close* (the emotional intensity of the relationship) and *behaving close* (the actual behavior that occurs within the relationship) (Aron, Aron, & Smollan, 1992). Both type of closeness have traditionally been studied by self-report data collected from participants themselves (Binder et al., 2012; Fu, 2007; Wellman & Wortley, 1990). However, this type of data collection is very time consuming and thus is limited in the level of detail that can be collected about specific interactions, the number of participants and social relationships that can be studied, and by participants' accuracy in recalling details of specific interactions with members of their social network (Bernard, Killworth, Kronenfeld, & Sailer, 1984; Boase & Ling, 2013).

More recently, is has become possible to measure the communication aspect of behaving close on an unprecedented level of detail, scale and accuracy based on big data sources such as email (Godoy-Lorite, Guimerà, & Sales-Pardo, 2016), Facebook (Burke, Kraut, & Marlow, 2011), Twitter (Grabowicz, Ramasco, Moro, Pujol, & Eguiluz, 2012) and mobile phone interactions (Blondel, Decuyper, & Krings, 2015; Saramäki & Moro, 2015). In these studies, Facebook interactions, tweets or mobile phone communication have been often been used as proxies for the underlying closeness of the social relationship (Bond et al., 2012; Gilbert & Karakalios, 2009), based on the assumption that more interaction indicates higher levels of closeness (S. G. B. Roberts & Dunbar, 2010). In studies based on mobile phone data, because mobile phone numbers are not typically publicly available, researchers have assumed that the presence of calls between two mobile users reflects a degree of personal relationship (Miritello et al., 2013; J-P Onnela, 2007). Further, the number or duration of calls is often taken as a direct indicator of 'tie strength' between two mobile users (Jo, Saramäki, Dunbar, & Kaski, 2014; Miritello et al., 2013; J-P Onnela, 2007; Palchykov, Kaski, Kertesz, Barabasi, & Dunbar, 2012; Palchykov, Kertész, Dunbar, & Kaski, 2013); for reviews, see (Blondel et al., 2015; Saramäki & Moro, 2015). In these cases, a (presumably linear) increase in mobile phone interaction is assumed to be associated with a linear increase in the strength of the social relationship. Thus a dyad who exchange 50 calls in a given time period is assumed to have a social tie that is 10 times as strong as a dyad who exchange 5 calls in same time period.

Further, in some cases, these measures of tie strength are subsequently used to model other properties of the communication network, such as community structure, network robustness or information flow (Blondel et al., 2015; J-P Onnela, 2007). Thus the accuracy of the conclusions being drawn from studies of large scale mobile datasets depends crucially on the number of calls being a reliable measure of the nature of the social relationship between mobile users. If the association between the number of calls and tie strength is weak or unreliable, then much more caution should be used when modelling social relationships and social structure based on mobile datasets (Wiese, Min, Hong, & Zimmerman, 2015). As Wiese et al. (2015, p. 765) caution: "The idea you can measure tie strength with relatively sparse data is tantalizing but dangerous". However, exactly what is meant by 'tie strength' is often left unspecified and empirical data showing a link between mobile communication and the strength of social relationships is limited, as in many studies only the records of mobile communication are available.

These limitations have resulted in calls for these large-scale studies based on data from a single channel (big data) to be supplemented with studies based on 'rich' or 'deep' data, which combine data across several different modes of communication and/or supplement digital data with information collected from the participants about the nature of their social relationships (Saramäki & Moro, 2015; Stopczynski et al., 2014; Wuchty, 2009). A small number of studies based on this rich data have allowed for the assumption of a link between the volume of mobile communication and tie strength to be empirically tested. For example, based on an 18-month study combining questionnaire and mobile data, Saramäki et al. (2014), showed that the number of phone calls explained 30-40% of variation in emotional closeness, and  10-20% of variation in the frequency of self-reported face-to-face contact. Further, Wiese et al. (2015) found that relying solely on call or text message communication to classify ties into 'strong' or 'weak' produced many errors, and concluded that communication frequency is a very noisy and imprecise indicator of tie strength. Thus mobile interaction appears to be significantly related to tie strength, but not in a very precise or reliable way.  Further, there is little understanding of how mobile interaction relates to other forms of behaving close in social relationships, such as face-to-face interaction.

In this study, we explore this issue using one of the largest and most comprehensive multi-channel smartphone datasets thus far collected, the Sensible DTU dataset (Stopczynski et al., 2014), which forms part of the Social Fabric project (www. http://socialfabric.ku.dk/). The Sensible DTU study had several waves of data collection (Stopczynski et al., 2014) and the participants in this study were 1,000 first year students from the Technical University of Denmark (DTU), who were provided with smartphones during the 2013/2014 academic year. The central research question we address in this paper is the relationship between two key channels of behaving close in social relationships – face-to-face interaction and mobile phone communication. In contrast with electronic communication which leaves a digital trace (D. Lazer et al., 2009), face-to-face interaction typically leaves no digital trace and is time consuming to observe directly or ask participants to record (Fu, 2007). However, Bluetooth sensors on mobile phones can record patterns of face-to-face interaction on a scale and level of detail not possible with either direct observation, or by asking participants directly (Eagle, Pentland, & Lazer, 2009). For example, using a combination of Bluetooth and mobile data, Eagle et al. (2009) were able to accurately infer 95% of self-reported friendships in a University population. However, this study did not directly examine the relationship between face-to-face communication (as measured by Bluetooth interactions) and mobile phone interaction. Further, how different aspects of face-to-face interaction, such as the number of interactions, the duration of interactions and the physical proximity during the interaction relate to patterns of mobile phone communication has not been systematically examined. The key goal of this paper is thus to explore how 'behaving close' face-to-face across different dimensions (number of meetings, duration of meetings, proximity during meetings) is associated with 'behaving close' over mobile phones. To the best of our knowledge, this is the first study to systematically explore the relationship between patterns of mobile communication and patterns of face-to-face interaction.

We focused on two key aspects of the larger Sensible DTU dataset (Stopczynski et al., 2014) to explore this issue. First, complete data on calls and text messages was recorded over the course of the study. Second, data on physical proximity was collected using Bluetooth sensors on the mobile phones. Different types of social relationships are based on different patterns of face-to-face interaction, with closer friends having both more frequent and longer interactions (Milardo, Johnson, & Huston, 1983; S. G. B. Roberts & Dunbar, 2010; Vlahovic, Roberts, & Dunbar, 2012; Wellman & Wortley, 1990) and interacting in closer physical proximity (Cristani et al., 2011; Goffman, 1963; Hall, 1966; Heshka & Nelson, 1972; Smith & Faig, 2014). Based on this literature, we developed three measures based on Bluetooth data: *meetings*, *duration* and *proximity*. A single Bluetooth connection between two phones (a *scan*) can imply a very short interaction between users A and B, but by combining these scans into cohesive *meetings* where the users are not separated for more than 15 minutes (sequential scans), we can identify more lengthy interactions between A and B. Second, we measured the *duration* of these meetings. Finally, we used the strength of the Bluetooth signal between the two phones to examine physical *proximity* between A and B during the interaction.

This study advances previous literature in in three key ways. First, whilst previous studies have explored the relationship between mobile phone communication and self-reported friendship (Eagle et al., 2009; Saramäki et al., 2014; Wiese et al., 2015), these studies did not investigate how mobile communication is related to face-to-face interaction. A significant relationship between these two variables would provide confidence in using mobile communication as a measure of tie strength. Second, existing studies measuring face-to-face interaction using proximity sensors have tended to simply use the raw number of scans as a measure of face-to-face interaction, rather than explore how further processing of the data in relation to the duration and proximity of interactions may be related to different types of social interactions. Finally, the study exemplifies the value of multi-channel data in understanding how we can use the digital trace of electronic communication to better understand social relationships. This underlines the importance of increasing not just the volume of data collected, in terms of number of participants or number of interactions, but also the number of channels over which data is collected (David Lazer, Kennedy, King, & Vespignani, 2014).

**Method**

**Dataset and participants**

We base our analyses on the Sensible DTU dataset (Stopczynski et al., 2014). Our analyses used records of calls and text messages between participants in the study over the course of 10 months during the 2013/2014 academic year. The dataset included information on the caller (the 'ego'), the receiver (the 'alter') and the time, date and duration of the call. For texts, the dataset included the sender, the receiver and the time and date the text was sent. Because previous research based on mobile datasets has tended to use the number of calls as an indicator of tie strength (J-P Onnela, 2007; Saramäki & Moro, 2015), we focused our main analysis on the number of calls between ego and alter. We report on parallel analysis using the duration of calls and number of text messages in the Electronic Supplementary Information (ESM). These phone interactions were then compared to the face-to-face meetings of the mobile users, as inferred from the Bluetooth sensors on the smartphones. The sensors were set to scan the surroundings (with an approximate range of 10m) every 5 minutes, recording any Bluetooth devices within range belonging to other participants in the Social Fabric study. The Bluetooth data included information on identity of the two users' mobile phones (ego and alter) and the time and date of the Bluetooth scan. We also extracted the signal strength of the Bluetooth signal (Received Signal Strength Indicator, RSSI). The RSSI measures the amount of power used to connect to other Bluetooth devices and as all the participants in the 2013/2014 cohort of the DTU study were issued with the same make and model of mobile phone (Stopczynski et al., 2014), the properties of the Bluetooth sensors and antennas were consistent across all participants. When measured across the same types of mobile phones, RSSI is very closely related to the physical distance between the two phones (Sekara & Lehmann, 2014). Thus in this study we used RSSI to measure the physical proximity of ego and alter during face-to-face interactions i.e. how close the two people were at the time of the Bluetooth scan.

The Sensible DTU dataset also contained information on participant characteristics (Stopczynski et al., 2014) and we included these as control variables in all analyses. We constructed social network matrices based on the gender of the dyad (male-male, female-female, male-female) and whether or not the two members of the dyad were in studying the same subject at DTU (study programme). Previous research relating mobile interaction to social relations has indicated the importance of both considering the location and circadian rhythms of mobile phone use (Aledavood et al., 2015; Eagle et al., 2009). Thus we explored whether the timing or location of face-to-face interactions was related to variance in the number of phone calls between dyads. Based on the study timetable for DTU students, we defined work hours as Monday-Friday 09:00-16:00 and non-work hours as any time outside this period (mornings, evenings and weekends). Further, we used a combination of GPS recordings and WiFi transmitters around the DTU campus to establish whether each Bluetooth interaction took place on campus or off campus (Sapiezynski, Stopczynski, Gatej, & Lehmann, 2015)

Throughout the study, some students dropped out of the university, dropped out of the study or simply stopped using their phones. To remove such cases from the dataset, only participants contributing both phone and Bluetooth data until the end of the tenth month were included in the dataset. Further, to avoid calls to voice mails, and students using their text messages inbox as reminders, phone interaction with participants' own numbers were removed. Finally, in order to maintain privacy, all participants were anonymized by giving them a unique identification number. Following this processing of the data, the final dataset contained 510 unique participants, with an age range from 19 to 46 (M = 21.4 year, SD = 2.6). In line with the student population at DTU, a large majority - 389 (78%) - of the participants were male.

Data collection, anonymization and storage was approved by the Danish Data Protection Agency and complies with local and EU regulations. All participants gave written informed consent electronically, with participants digitally signing the consent form with their university credentials. For full details of how ethical and privacy issues were addressed throughout the course of the study, see Stopczynski et al. (2014).

**Data processing and statistical analysis**

We used social network analysis to examine the relationship between mobile phone interactions and face-to-face interactions. For the social network matrix based on phone interaction, each cell represented the amount of phone interaction (number of calls) between ego and alter over the 10 month study period (Figure 1).

INSERT FIGURE 1 HERE

For the Bluetooth data, each of the 510 participants' phones scanned their surroundings every 5 minutes for a period of 10 months, generating a very large amount of Bluetooth data per participant. Due to the novelty of this data type, there is little systematic knowledge available on how to organize and filter the Bluetooth data to remove the noise and relate the Bluetooth data to meaningful social interactions. Thus, as a first step we examined how the Bluetooth interactions related to phone interactions for all of the dyads in the dataset (Table 1). In total, there were 1,046 dyads which had some phone interaction and some Bluetooth interaction during the study period and this is the sample we used in all our analyses. There were over 100,000 dyads that did not have any phone communication, but did have some Bluetooth interaction. Our key focus in this study is examining the extent to which variance in mobile phone communication can be explained by variance in Bluetooth interaction, as in many mobile datasets the only information available is the mobile phone communication between users  (Blondel et al., 2015; Saramäki & Moro, 2015). We thus did not use in the analyses the large number of dyads which had some Bluetooth interaction but no phone interaction over the study period. As these dyads had no phone interaction, there was no variance in the number of calls that could be explained based on Bluetooth interaction.

INSERT TABLE 1 HERE

Based on this dataset of 1,046 dyads, we explored how different Bluetooth matrices were related to mobile communication. The first network is based on the number of *raw scans* detected between two participants' phones over the 10-month period. We then aggregated the number of raw scans into *meetings*, consisting of consecutive scans where the two phones are not separated for more than 15 minutes. If the two phones are separated for more than 15 minutes, this is then counted as a new meeting. Additionally, by counting the number of scans in each meeting we are able to measure the *meeting duration* in seconds. Based on this, we calculated the *mean meeting duration* per dyad. Finally, we also developed a measure of the *mean proximity* per dyad, based on the mean RSSI value between two users' phone during all scans between those phones. In line with previous research (Saramäki et al., 2014) the number of calls, duration of calls and number of texts were log transformed, to reduce the effect of outliers on the results, and make the relationship between calls and the Bluetooth variables more linear and thus suitable for regression analyses. Similarly, the number of scans and number of meetings were also log transformed. The mean meeting duration and mean proximity were not log transformed, as these were based on means rather than raw data and thus had fewer outliers.

In order to examine the relationship between social networks based on phone interaction and networks based on face-to-face interaction, we used network regression techniques. Multiple Regression Quadratic Assignment Procedure (MRQAP) is a form of network regression that allows for a dependent matrix (phone calls) to be regressed against multiple independent explanatory matrices (face-to-face interaction and control variables). Firstly, a standard ordinary least squares regression is run on the data and an $R^2$ and coefficients for the independent variables are calculated. Then, the rows and columns of the dependent matrix are randomly permutated a large number of times (in our analyses 100 times) whilst retaining the topology of the network and the regression statistics recalculated. This creates a matrix-specific distribution against which observed $R^2$ and coefficients for the actual data can be compared to determine significance. This permutation approach takes into account the autocorrelation inherent to social network data, in that each dyadic data point is not independent, meaning standard statistical tests cannot be used. We used Dekker's Double Semi-Partialling procedure which is more robust against mutlicollinearity in the data than standard MRQAP regression (Dekker, Krackhardt, & Snijders, 2007). Since our key independent matrices are all based on Bluetooth data, we also checked for multicollinearity by running a standard least squares regression and calculating the Variance Inflation Factors (VIF). VIF values of above 10 indicate potential collinearity issues (Field, 2013). We used the Akaike Information Criterion (AIC) to assess relative model fit. AIC penalises adding additional parameters to the model, and thus gives an indication of the most parsimonious model that best fits that data.    All analyses were carried out using the using the *netlm* function in the R package statnet/sna (Butts, 2014).

## Results

### Descriptive statistics

The average dyad called each other 17 times in the 10 month period and had over 3,000 Bluetooth scans, which translates into approximately 135 hours of interaction time (Table 2). Each dyad had a mean of 114 meetings, with a mean meeting duration of over an hour. The mean RSSI for each dyad was 18, which translates to a proximity of around 2m. For all of these values, there is a large standard deviation, a common finding in behavioural data based on digital traces (Miritello et al., 2013; Saramäki et al., 2014).

INSERT TABLE 2 HERE

Little is known about the nature of the dyads represented by different patterns of Bluetooth data (Sekara & Lehmann, 2014). Thus in an exploratory analysis, we examined the 'relational types' amongst our dyads, based on patterns of face-to-face interaction, programme of study and location of scans.  Of our 1,046 dyads, 67% are registered in the same study program, showing that a large number of relations (around a third) were not solely based on studying the same subject.  As may be expected, relations formed across study programmes tend to meet outside of work hours, whilst same relations formed within study programmes are more likely to during work hours (ESM, Figure S1). However, the majority of same study relations meet both during and outside of working hours. Further, the largest single group of dyads (28%) have over 100 meetings and meet both during work and non-work hours (ESM, Figure S2). Overall, despite all studying at the same University, the majority our dyads have over 30 meetings and do not primarily meet during working hours. This suggests that the Bluetooth scans reflect genuine social relationships and do not simply represent dyads who happen to be studying the same programme and sitting in close proximity during lectures or seminars, or transient relations who only meet a small number of times over the study period.

**Statistical analysis**

The first analysis examined the pairwise Pearson correlations between all variables (Table 3). There are statistically significant correlations between the number of calls and all measures of face-to-face activity. None of the correlations between the independent variables are over 0.7, suggesting  that these variables can be entered into a single regression model without problems relating to multicollinearity (Field, 2013).

INSERT TABLE 3 HERE

Table 4 presents the results from the MRQAP network regression, with number of calls as the dependent network. Our baseline model contains only the control variables – none of these variables are significant and these variables explain less than 1% of the variance in the number of phone calls. In Model 1, the number of scans between a dyad is significantly related to the number of calls between that dyad, with the regression model accounting for 21% of variation in number of calls. When these scans are aggregated into meetings (Model 2), both the number and duration of meetings significantly predicts the number of calls. Thus simply aggregating the scans into longer meetings increases the amount of variance explained to 24% and reduces the Akaike Information Criterion (AIC), indicating better relative model fit. In Model 3, the mean proximity of the dyad during meetings was entered separately and this model explains more variance in calls (27%) than Model 2, which included the number and mean duration of meetings. Finally, Model 4 includes all the variables relating to face-to-face interaction – the number of meetings, mean duration and mean proximity. This final model explains 36% of variance in the number of calls between dyads and has the lowest AIC despite having the largest number of parameters, indicating the best relative model fit. The standardised coefficients indicate that proximity during meetings is the single most important variable in explaining variance in phone calls. The non-standardised coefficients indicate that a proportional increase of 25 meetings translates into 1 extra phone call, and every increase of the average proximity by 10, measured as RSSI, relates to 1 extra phone call[1]. While the duration is significant, the effect is not large, with an increase in the average meeting duration of 500 minutes translating into just 1 extra phone call. In Models 1-4, study programme was negatively related to calls, indicating that dyads who were in different study programmes made more calls. For all models, the VIF values did not indicate any collinaerity issues, with the largest VIF between the primary variables measuring under 1.7.

INSERT TABLE 4 HERE

        In additional analyses, we explored how the results would change if the dependent network was based on the duration of calls or number of text messages, rather than the number

---

[1] Since proximity (RSSI) is measured on an exponential scale, the meaning of an increase in RSSI of 10 changes depending on the absolute RSSI value. However for the average dyad with a mean of around 18 (~2 metres), an increase of 10 would translate into roughly 1m.

of calls (Supplementary Information, Tables S1, S2). Briefly, the results were similar for both text messages (Supplementary Information, Table S3) and duration of calls (Supplementary Information, Table S4), with physical proximity during meetings the single strongest predictor and the combined model containing number, duration and proximity of meetings explaining the greatest percentage of variance. We also examined whether information about the timing or location of face-to-face interactions helped explain variance in the number of calls between dyads. Splitting meetings into those that took place during working and non-working hours reduced the fit of the model, explaining less variance in calls compared to Model 4, which did not include time of calls (Supplementary Information, Table S5). Further, meetings that took place during working hours were a stronger predictor of calls than meetings that took place during non-working hours. Similarly, including information as to whether the meeting took place on campus or off campus also reduced the fit of the model compared to Model 4, which did not include location (Supplementary Information, Table S6).

### Discussion

In this study, we used one of the largest and most comprehensive multi-channel datasets yet collected to examine the relationship between mobile communication and face-to-face interaction. To the best of our knowledge, this is the first study to systematically examine the association between these two important aspects of behaving close in social relationships. There were three key findings. First, over a 10 month period, there was a significant relationship between the number of phone calls between a dyad (people A and B) and the number of face-to-face meetings between the dyad. In the final regression model, 36% of the variance in the number of phone calls between the dyad could be explained using variables relating to the face-to-face meetings and the control variables. Second, the physical proximity between people A and B during the face-to-face interaction was the single most important predictor of the number of phone calls between the dyad. Third, processing Bluetooth data in ways related to key features of social interactions – longer meetings and proximity during these meetings – increased the amount of variance in phone calls we were able to explain, as compared to simply using the raw number of Bluetooth scans between two phones. This pattern of results was similar for number of phone calls, duration of phone calls and number of text messages. Overall, face-to-face communication is significantly related to mobile communication, and physical proximity during face-to-face interaction is a more important predictor of the number of calls between dyads than the number or duration of meetings.

Closeness in social relationships can be characterized by two components – feeling close and behaving close (Aron et al., 1992). Our interpretation of the results is that people in social relationships 'behave close' across a number of different modes of communication, including both mobile phone calls and face-to-face interaction. Thus on average, dyads that had a greater number of calls also had more face-to-face meetings. This conceptualization of relationships mirrors theories around media multiplexity of communication, where rates of communication across different channels (face-to-face, mobile, email) are closely related to each other (Haythornthwaite, 2005).

A key novel finding in this study was the significant relationship between mobile communication and physical proximity. Previous work has simply used Bluetooth scans as an indication that two people were within 10m of each other (Aharony, Pan, Ip, Khayal, & Pentland, 2011; Eagle et al., 2009). However, the nature of a face-to-face interaction may vary greatly according to whether the interaction happens within 10m, 5m or 1m, with closer proximity associated with stronger social relationships (Cristani et al., 2011; Heshka & Nelson, 1972). Because all the participants were using the same mobile handsets (Sekara & Lehmann, 2014), in this study for the first time we were able to relate proximity to mobile communication. Students build and maintain a wide variety of social relationships over the course of their first year at University (S. B. G. Roberts & Dunbar, 2015). Our results show that physical proximity during face-to-face interactions appears to be one important feature that can be used to distinguish not just between strangers and friends (Heshka & Nelson, 1972), but within the wide range of social relationships people maintain with others, from casual acquaintances to close friends. Consciously or sub-consciously, people place themselves physically closer to others with whom they also have more mobile phone interaction. Given the widespread use of smartphones (and thus Bluetooth sensors), future work can explore in more detail how interpersonal distance changes as the social relationship develops (Birkelund, Bornakke, & Glavind, 2015) and how proximity may differ according to age, gender and cultural background (Cristani et al., 2011; Heshka & Nelson, 1972).

Other factors related to face-to-face interactions that we expected to be important in explaining variance in phone calls did not appear to play a significant role in this sample. In contrast with previous research (Aledavood et al., 2015; Eagle et al., 2009), there was no indication that face-to-face meetings outside of work hours, or off campus, were more predictive of the number of phone calls between a dyad than face-to-face meetings during work hours or on campus. This picture is reinforced by the finding that even among students studying the same programme, most dyads did not meet exclusively during working hours. Students appear to have a more fluid socialising and working style across time and space than typical nine-to-five office workers. Students may socialise face-to-face on campus during working hours, as they are not in classes for the whole of the working day, and work off campus during 'non-working' hours. In contrast, if our sample had been the academic staff of the university, rather than students, we would expect a clearer division between work and leisure time (both in time and space) and thus a greater predictive power of these variables to explain variance in calls. Thus issues around sampling and the generalizability of research findings are as relevant in interpreting 'big data' sources as in more traditional studies reliant on questionnaires (boyd & Crawford, 2012; David Lazer et al., 2014).

More broadly, these results raise issues about using mobile interaction as a precise and accurate indicator of tie strength (Jo et al., 2014; Miritello et al., 2013; J.-P. Onnela et al., 2007; Palchykov et al., 2012; Palchykov et al., 2013; Palla, Barabasi, & Vicsek, 2007). One way to interpret our results would be to focus on the fact that face-to-face meetings (together with the control variables) explain around 36% of variance in mobile phone calls. However, another interpretation would be to focus on the 64% of variance in phone calls not explained by the variables in our models. Whilst explaining 36% of the variance in the dependent variable may be considered high for two separate constructs (e.g. phone interaction and well-being), it could be considered low for variables measuring the same underlying aspect of social relationships – behaving close. Thus mobile calls appear to be a rather noisy and imprecise signal of 'behaving close', in that many dyads will have a high number of phone calls with low frequencies of face-to-face interaction, and vice versa.

Further, this study may overestimate how much variance in calls is explained by face-to-face interaction, as we excluded all dyads which had no mobile phone communication over the study period. For a very large number of these dyads, there was some degree of face-to-face interaction, suggesting some type of social relationship. If only mobile data was available, these 'Bluetooth only' dyads would be missed entirely, and thus the communication network based on mobile phone interaction alone would an inaccurate representation of the underlying social network. This may have important implications for the accuracy of conclusions based on mobile data alone. For example, if the true communication network is much denser than the reported mobile network due to many important ties missing from the mobile network, the pattern of information flow through the network may be quite different than that predicted based purely on a mobile dataset (Blondel et al., 2015; J-P Onnela, 2007). Examining the nature of these dyads with face-to-face interaction but no mobile communication is beyond the scope of the current study, but future research is needed into what type of social ties mobile data and other forms of electronic communication captures, and what type these datasets miss (Wiese et al., 2015).

In conclusion, big data sources do have the potential to offer new insights into social relationships (D. Lazer et al., 2009), particularly when these data are collected across multiple channels over a long time period (Aharony et al., 2011; Stopczynski et al., 2014). The detailed data reported in his study on face-to-face interactions and mobile communication would not have been possible to collect if reliant on traditional questionnaire or interview methods of data collection. However, even with detailed information about the pattern of face-to-face interaction over ten months, less than 40% of variation in the number of phone calls between dyads was explained by the regression models. Future work should focus on improving our understanding of the relationship between the different channels of communication people use to maintain their relationships (face-to-face interaction, mobile communication, communication over the internet) and the two key components of social relationships - behaving close and feeling close (Mastrandrea, Fournet, & Barrat, 2015). This research should take into account the growing trend for mobile users to use messaging applications (such as WhatsApp or Facebook messenger) in place of text messages, meaning records of calls and texts may become less accurate at reflecting the actual communication between a dyad (Saramäki & Moro, 2015; Wiese et al., 2015). Mobile numbers are not typically publicly available, so the presence of mobile interaction between two people can be used to infer that these people have some sort of social relationship. Further, by definition the amount or duration of mobile interaction provides an indication of the time invested in that relationship (Miritello et al., 2013). However, until the relationship between the digital record of behaviour and the nature of the underlying social relationship is better understood, in our view mobile communication should be considered as a useful but noisy measure of tie strength with some significant limitations (boyd & Crawford, 2012; Wiese et al., 2015). Caution should therefore be used when making inferences about tie strength in social relationships based solely on mobile communication patterns.

Table 1. The number of dyads with phone interaction and/or Bluetooth interaction over the 10 month study period. This table only includes dyads that met the data quality requirements described in the Methods section.

| | | Bluetooth interaction | |
|---|---|---|---|
| | | Yes | No |
| Phone interaction | Yes | 1,046 | 488 |
| | No | 138,026 | n/a |

Table 2. Descriptive data for phone communication and Bluetooth interactions between dyads. RSSI is the Received Signal Strength Indicator, which measures the strength of the Bluetooth signal between two mobile phones.

|                               | Mean | SD   |
|-------------------------------|------|------|
| Number of calls               | 17   | 43   |
| Number of scans               | 3238 | 5539 |
| Number of meetings            | 114  | 117  |
| Duration of meetings (mins)   | 82   | 59   |
| Proximity of meetings (RSSI)  | 18   | 6    |

Table 3. Bivariate Pearson's correlations between mobile phone communication and Bluetooth interactions. P values calculated using vector permutation test.

| | Scans | Meetings | Mean meeting duration | Mean meeting proximity |
|---|---|---|---|---|
| Calls (log) | 0.409*** | 0.328*** | 0.421*** | 0.518*** |
| Scans (log) | | 0.902*** | 0.604*** | 0.474*** |
| Meetings (log) | | | 0.256*** | 0.204*** |
| Mean meeting duration | | | | 0.591*** |

Note: ***$p < 0.001$

Table 4. MRQAP network regression, predicting the logged number of mobile phone calls between dyads, based on face-to-face interaction and control variables (gender, programme of study). Table shows regression coefficients, with standardised coefficients in brackets.

| Predictor | Baseline | Model 1 | Model 2 | Model 3 | Model 4 |
|---|---|---|---|---|---|
| Adjusted $R^2$ | 0.006 | 0.211 | 0.243 | 0.271 | 0.357 |
| AIC | 3664 | 3465 | 3348 | 3307 | 3183 |
| Model significance | * | *** | *** | *** | *** |
| Scans (log) | | 0.439 | | | |
| | | (0.449)*** | | | |
| Meetings (log) | | | 0.377 | | 0.373 |
| | | | (0.283)*** | | (0.287)*** |
| Mean meeting duration | | | 0.008 | | 0.002 |
| | | | (0.352)*** | | (0.099)** |
| Mean meeting proximity | | | | 0.135 | 0.110 |
| | | | | (0.531)*** | (0.432)*** |
| Study programme (same/different) | 0.111 | -0.387 | -0.243 | -0.187 | -0.455 |
| | (0.040) | (-0.123)*** | (-0.086)** | (-0.066)* | (-0.160)*** |
| Male-male | 0.058 | 0.302 | 0.487 | 0.071 | 0.337 |
| | (0.020) | (0.007) | (0.174) | (0.025) | (0.121) |
| Female-female | -0.064 | 0.336 | 0.524 | -0.099 | 0.271 |
| | (-0.013) | (0.006) | (0.113) | (-0.020) | (0.061) |
| Female-male | -0.205 | 0.149 | 0.338 | -0.009 | 0.292 |
| | (-0.068) | (-0.058) | (0.111) | (0.003) | (0.096) |

Notes: *$p < 0.05$, **$p < 0.01$, ***$p < 0.001$. AIC is the Akaike Information Criterion

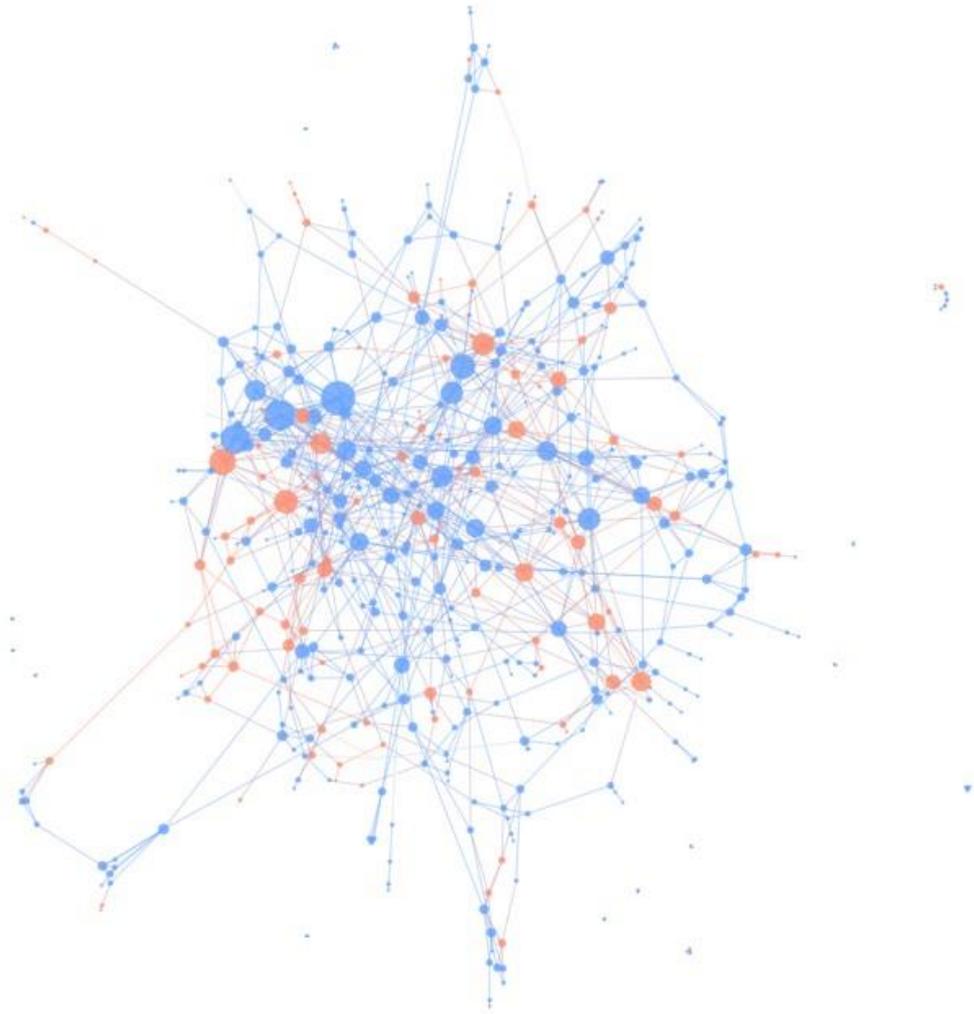

Figure 1. Network based on the number of mobile phone calls between dyads. Red nodes are

females, blue nodes are males. The size of each node reflects the average weighted degree –

the average number of phone calls to other participants.